\documentstyle[aps,prl,eqsecnum,preprint,tighten]{revtex}

\begin{document}
\def\bq{\begin{equation}}
\def\eq{\end{equation}}
\title
{ LEVEL STATISTICS INSIDE THE VORTEX OF A SUPERCONDUCTOR
  AND SYMPLECTIC RANDOM MATRIX THEORY IN AN EXTERNAL SOURCE
 }
\author
{ E. Br{\'e}zin$^{1}$, S. Hikami$^{2}$ and A. I. Larkin$^{c),d)}$} 
\address
{$^{1}$ Laboratoire de Physique Th{\'e}orique, Ecole Normale
Sup{\'e}rieure, 
24 rue Lhomond 75231, Paris Cedex 05, France{\footnote{
Unit{\'e} propre du centre national de la Recherche Scientifique, Associ{\'e}e
{\`a} l'Ecole Normale Sup{\'e}rieure et {\`a} l'Universit{\'e} de Paris-Sud} 
}\\
$^{2}$ Department of Pure and Applied Sciences, University of Tokyo\\
{Meguro-ku, Komaba, Tokyo 153, Japan}\\
{$^{c)}$ Theoretical Physics Institute, University of Minnesota,
Minneapolis, Minnesota 55455, USA}\\
{$^{d)}$ Landau Institute for Theoretical Physics, 117334 Moscow, Russia}\\}
\maketitle
\begin{abstract}
 In the core of the vortex of a superconductor, energy levels appear inside
the gap. We discuss here  through a random matrix approach
how these levels are broadened
by  impurities. It is first shown that the level statistics is governed
by an ensemble consisting of a symplectic random potential added to a
non-random matrix. A generalization of previous work on the unitary
ensemble in the presence of an external
source (which relied on the Itzykson-Zuber integral) is discussed for this
symplectic case through the formalism introduced by
Harish-Chandra and Duistermaat-Heckman. This leads to explicit formulae for
the density of states and for the correlation functions, which describe the
 cross-over from the clean to the dirty limits.
\end{abstract}
\vskip 5mm
\section{INTRODUCTION}
\vskip 3mm

The energy levels inside a vortex of a superconductor
have been characterized  long  ago \cite{CDM}, but recent studies have
dealt with the broadening of these levels by impurities. The density of
states has been
computed both in the clean limit
and for the dirty case.

In this article, we discuss the crossover between these two  limits. The
impurity potential is handled through a  random matrix theory. However
the matrix elements between different energy levels due to the impurities
are strongly correlated,
and therefore one is far from a usual Wigner-Dyson random matrix theory.

It is shown below that the crossover is characterized in this case by a
random symplectic matrix coupled to
an external non-random source matrix. For handling this problem we develop
a technique which generalizes
 earlier work on the unitary ensemble in which one considered  the cross-over
from a deterministic hamiltonian to a fully random hermitian hamiltonian
\cite{BH1,BHZ,BH2,BH3}.
 In the hermitian case, the formalism relied on an integral over the
unitary group due to Harish-Chandra \cite{Harish-Chandra} and
rederived in the context of random matrix theory by Itzykson-Zuber
\cite{Itzykson}. All the correlation functions for the energy levels were
then found explicitely.
For the vortex problem considered here the perturbation due to impurities
is a matrix with a symplectic structure,
which we treat as a random potential added to the unperturbed  diagonal
matrix consisting of the regularly spaced energy levels.
It is shown below that for this problem a
 similar integral over the symplectic group, considered  by Harish-Chandra
\cite{Harish-Chandra}, and more recently  generalized by
 Duistermaat and Heckman \cite{Duistermaat} leads also to explicit formulae.
The crossover from the clean spectrum
to the dirty limit follows from this formalism.
We will show how this general behavior for the crossover is relevant to the
actual problem of the excitation energy spectrum in a superconducting
vortex.

\section{SYMPLECTIC STRUCTURE OF THE 
PERTURBATION FOR THE ENERGY LEVELS INSIDE A VORTEX}

The energy eigenvalues and eigenfunctions of a quasi-particle
inside the vortex of a superconductor,
were obtained
 long ago  \cite{CDM}.
The two component wave functions of excitations $\hat \psi = $
$( \psi_1,\psi_2 )$ in a superconductor satifies the Bogolubov-de Gennes
equation,
\bq\label{2.1}
[ \sigma_z (H_0 + V_{imp}) + \sigma_x {\rm Re}\Delta(r) + \sigma_y
{\rm Im} \Delta(r)] \hat\psi = E \hat\psi
\eq
where $H_0 = p^2/2m - E_F$,  and $\Delta(r)$ is
a gap order parameter. The  impurity potential $V_{imp}$ is a sum of
short range scattering sites $r_i$:
\bq\label{2.2}
V_{imp}(r) = \sum_i V_i\delta(r - r_i)
\eq
In the absence of impurity, the  Schrodinger equation
(\ref{2.1})  leads to a spectrum of equidistant states  $E_n^0$
\bq\label{2.3}
E_n^0 = - \hbar \omega_0(n - {1\over{2}})
\eq
where $n$ is an integer defining the quantized angular momentum of
 the vortex,
and  $\hbar \omega_0$ = $\Delta^2/E_F$. We assume that
the spacing $\hbar \omega_0$  between these levels is much smaller than the
gap  $\Delta$,
and consequently there are many excitation levels in the vortex core.

Using the explicit wave functions for the unperturbed eigenstates, one
finds that the matrix elements of the interaction
due to the $i$-th impurity are
\bq\label{2.7}
A_{nm}^i = {V_i e^{- 2 K(r_i)}\over{\lambda_F \xi}}
e^{i ( m - n)\phi_i}[ J_n(k_F r_i) J_m(k_F r_i)
- J_{n - 1} (k_F r_i) J_{m - 1}(k_F r_i)]
\eq
where $(r_i,\phi_i)$ is the position of the i-th impurity in  polar
coordinates.  The indices  $n$ and $m$ run from $-N$ to $+ N$ where
$N \sim \Delta/\omega_0 \sim E_F/\Delta$
 is assumed  to be a large number.

The scattering time $\tau$, given by $1/\tau = 2 n_i V_i^2 m$,
where $n_i={N_i}/{\xi^2}$ is the density of impurities in the vortex ;
$N_i$ the total of impurities in the cross-section of a vortex of radius
$\xi$.\\
Three different situations may occur:
a) a dirty limit $\Delta << 1/\tau << E_F$, b) a clean limit
$\omega_0 < 1/\tau << \Delta$, c) a superclean case
$1/\tau << \omega_0 \sim \Delta^2/E_F$. However, in
 \cite{KL1,LO} it has been shown that the level statistics depends
not only on the parameter $\tau$ but also on the number of impurities $N_i$
. This is clear in the superclean case.
Then the density of states has  narrow peaks centered around  the
unperturbed energy levels
 (\ref{2.3}). The form of these peaks is Gaussian when
the number of impurities is large ; however as shown in \cite{KL2,LO} these
peaks have
 a different form
 when there is only a single impurity.

In the clean case $\omega_0 < 1/\tau < \Delta$, the level
statistics also depends upon the number of impurity $N_i$.
If $N_i \rightarrow \infty$ and if the random potential is a
white noise, it was shown in  \cite{SKF} that one can apply standard random
matrix theory.
However if the number of impurities $N_i$ is smaller than
a certain number $N_{ic}$ ($N_{ic} > (E_F/\Delta)^{1/2}$),
Koulakov and Larkin \cite{KL1} have found that the spectrum is of a
different type.

In the present article, we have defined the dirty and the clean limits
by $N_i > N_{ic}$, and
 $ N_i < N_{ic}$ respectively. (These definitions differ slightly  from the
usual
 ones, quoted hereabove.)

When there is only one impurity in the vortex core, with
a radius $\xi$, the impurity  is located at a position $r_i \sim \xi$.
We  are thus in a limit in which $k_F r_i >> 1$, since the correlation
length is
$\xi \sim \hbar^2 k_F/m\Delta$, and we have thus $k_F \xi \sim
k_F^2/\Delta \sim E_F/\Delta >> 1$.
Therefore, in the clean limit, this condition
is always satisfied. In the dirty limit, the number of
 impurities increase and the typical value of the
 quantity $k_F r_i$ is reduced.

In that limit the Bessel functions in (\ref{2.7}) may be replaced by their
asymptotic expansions and the
matrix element $A_{nm}^i$,  produced by the scattering site $i$ between two
states $n$ and $m$, is given for $k_Fr_i\gg 1$ by
\bq\label{2.11}
A_{nm}^i \simeq {\tilde C} e^{i (m - n)\phi_i} \sin( 2 k_F r_i - {n +
m\over{2}}
\pi )
\eq
where ${\tilde C} = {V_i e^{- 2K(r_i)}\over{\lambda_F \xi}}
({2\over{\pi k_F r_i}})$.

The corresponding secular equation for the clean case was solved by
Koulakov and Larkin\cite{KL1} who have found that, after averaging over the
location of the impurity,
the density of states $\rho(E)$ was given by
\bq\label{2.12}
\rho(E) = {2\over{\omega_0}}
\sin^2 ({\pi E\over{\omega_0}})
\eq
(normalized to one after  integration on the interval
$0 < E < \omega_0$).\\
 In the dirty limit, $k_F r_i$
is not as large and it is necessary to take into account the sub-asymptotic
behavior of the Bessel functions, and to
sum over the impurities.
Since the matrix elements $A_{nm}^i$ have an oscillatory behavior, it is
natural to
regard them in the dirty limit as the elements of a  random matrix. However
it is necessary to keep some of the structure exhibited
by the clean limit (\ref{2.11}) in mind.

In the dirty limit or
in the moderate clean case, a random matrix theory has already been considered
for the superconductor-normal
interface, and  for a quantum dot with a superconductor boundary.
\cite{AZ,AST}. The excitation levels of the
superconductor vortex have also been analysed in this way \cite{SKF}.
The random matrix ensemble, for such cases,  has been
suggested to be invariant by the symplectic group $Sp(N)$.

If the full Hamiltonian is treated as a random matrix in a symplectic
ensemble $Sp(n)$ , which is valid  in the extreme dirty limit, the
 density of states
has been conjectured  \cite{AZ,SKF} to be
\bq\label{2.13}
\rho(E) = 1 - {\sin ({ \pi E \over{\omega_0}})\over({\pi E\over{\omega_0}})}
\eq
This differs markedly from the clean result (\ref{2.12}).\\
  It is therefore challenging  to find the crossover behavior
between these two limits, and it might even  be important for the
transport problems associated with the excitations of quasi-particles
inside the vortex core \cite{KL2}. In order to achieve our goal, instead of
treating the full Hamiltonian as random,
 we consider here only the scattering by impurities
as a random matrix but not the
full Ha	miltonian.\\
In order to motivate the use of the symplectic structure in this problem
it is instructive to examine  the form of the  matrix
$A_{nm}^i$.
For  a given impurity $i$, let $A$ be the matrix whose elements are
 $<n|A|m>=A_{nm}^i  $.
The state index $|m>$ varies over $2 N$ levels. When $N = 2$, for example,
the states  $|m>$ are
 $|-1>,|0>,|1>,|2>$. Since $E_n^0 = -
 \omega_0( n - {1\over{2}})$, we have
 $E_0 = -E_1 = {1\over{2}}\omega_0$ and $E_2 = - E_{-1}
= -{3\over{2}}\omega_0$. Labelling the lines and rows of the matrix $A$ in
the order   $|0>,|2>,|1>,|-1>$,
in order to split it into even-even, even-odd, odd-odd $2\times 2$ blocks,
we write
\bq\label{2.14}
A = \left(\matrix{ <0|A|0>& <0|A|2> & <0|A|1> & <0|A|-1> \cr
<2|A|0>&<2|A|2>&<2|A|1>&<2|A|-1>\cr
<1|A|0>&<1|A|2>&<1|A|1>&<1|A|-1>\cr
<-1|A|0>&<-1|A|2>&<-1|A|1>&<-1|A|-1>
}\right)
\eq
Since $<0|A|2> = A_{02}^i = A_{20}^i= <2|A|0>$,
and $<2|A|1> = <0|A|-1>$, etc. from (\ref{2.7}) (we have used the
$J_{-n}(x) = (-1)^n J_{n}(x)$), it is easy to see that
the matrix $A$ has, for arbitrary $N$, the structure
\bq\label{2.15}
A = \left(\matrix{a & b\cr b^{*}& - a^{T}}\right)
\eq
where  $a$ is  an $N\times N$ hermitian matrix $a^{\dag} = a$, and $b$ is an
$N\times N$ symmetric
complex matrix $b^{T} = b$ ($b^{T}$ is  the transpose of $b$).

The number of degrees of freedom in the random matrix  $A$ is thus
$2 N^2 + N$, ( $N^2$ for the hermitian matrix $a$ and $N^2 + N$
for the complex symmetric matrix $b$).
The number of  generators of the symplectic $Sp(N)$ group
is indeed  also $N(2 N + 1)$.

The symplectic structure of $A$ is exhibited by the algebraic relation
\bq\label{2.17}
 A^{T} J + J A = 0
\eq
where $J$ is
\bq\label{2.18}
 J = \left(\matrix{ 0 & 1\cr
-1 & 0} \right).
\eq
This structure implies that the eigenvalues are real and pairwise opposite,
giving a chiral structure to the eigenvalues.
The Lie algebra $X$ is diagonalized by the symplectic group $G \in Sp(N)$.
For example, for $N = 2$, we have
\bq\label{2.19}
X = G^{\dag} \left(\matrix{\lambda_1 & & & \cr
                                     &\lambda_2& & \cr
                                     & & -\lambda_1& \cr
                                     & & &-\lambda_2}\right) G
\eq

If we mutiply the matrix $A$ by $i
=\sqrt{-1}$ one recovers  an element of the Lie algebra of the group $Sp(N)$.
If we consider now that  the position $(\phi_i,r_i)$ of the impurity is
random,  it is natural to
take $<n|A|m>$ in this $Sp(N)$ invariant
random ensemble.

For instance when there is only one impurity (i = 1), the matrix $A$ has a
periodic
structure, with alternating signs for each 2 by 2 block. For example,
in the case of N = 4, we have for the full Hamiltonian $H =  E^0 + A$,
($E^0$ is a diagonal matrix)
\bq\label{2.20}
A = \left(\matrix{ s & - s & - c & c\cr
                 - s& s & c & - c\cr
                 - c & c & - s & s\cr
                 c & - c & s & - s}\right)
\eq
where $ s = \tilde C \sin(2 k_F a)$ and $c = \tilde C \cos( 2 k_F a)$.
( In the single impurity case the phase $\phi$ can be
set equal to zero by a choice of gauge condition ).
This even-odd structure has been studied earlier in \cite{LO,KL1}.
The periodic structure becomes clearer if we
write the matrix $A = (A_{nm})$, (\ref{2.7}) for N=2 with
the approximation  (\ref{2.11}),

\bq\label{2.21}
A = \left(\matrix{s & c & -s & -c\cr
                  c & -s & -c & s\cr
                  -s & -c & s & c\cr
                  -c & s & c & -s}\right)
\eq
in which rows and lines are in the order
$|2>,|1>,|0>,|-1>$.
The determinant of this matrix is a kind of Toeplitz determinant (constant
along parallel to the anti-diagonal).
We have omitted the diagonal part $E_n^0$.
It is clear that the eigenvalues are periodic, depending on the
odd-even parity of $n$.
 When we consider several impurities, a sum over impurities $\sum_i
A_{nm}^i$ should be taken. In the clean limit,
we replace the quantities $s$ and $c$ by
$s = \sum_i {\tilde C} \sin (2 k_F r_i)$ and $c = \sum_i {\tilde C}
 \cos (2 k_F r_i)$.
The random average over the positions
of these impurities gives the density of state (\ref{2.13}) \cite{KL1}.

\section{ITZYKSON-ZUBER INTEGRAL AND ITS EXTENSION}

  In the previous section, we have discussed how we are led to study the
crossover
  from the clean case to the dirty case.
  We now divide the matrix $\sum_{i=1}^{N_i}A_{nm}^i$ into two parts.
  One part, denoted by $H_0$, corresponds to the clean  case, namely
  the matrix has a periodic structure as in (\ref{2.20}), like a Toeplitz
  matrix, but with a sum over $i$.
  The other part is the remaining difference between the matrix
$\sum_{i=1}^{N_i}A_{nm}^i$ and
  $H_0$, which is considered as a random matrix $V$.
  Thus we have
  \bq\label{3.1}
    H = H_0 + V
  \eq
  The matrices $H_0$ and $V$ are both symplectic, represented by the
  symplectic group Lie algebra $Sp(N)$, which has the form (\ref{2.15}).

   In the analogous, but simpler, problem of a unitary  invariant random
perturbation,
 ($H_0$ and $V$ were then both complex Hermitian matrices), a technique to
study
   the crossover from the non-stochastic $H_0$ to a pure random matrix $V$
has been developped earlier \cite{BH1,BHZ,BH2}.
   Here we have to consider the symplectic case. As in the unitary case,
   we have to integrate over the Lie group  which diagonalizes the
   matrix $H$.
   In the unitary case, the first step, the integration over the unitary group,
 was introduced in the study of two coupled random matrices by Itzykson and
Zuber.
\cite{Itzykson}.
   We now summarize the formulae of integration over Lie Groups, which are
   useful for the subsequent discussion. We assume that the probability
distribution of
   $V$ is a Gaussian, and that $H_0$ is a fixed non-random matrix. [It is
possible to
   generalize it to some non-Gaussian distributions for $V$ ; for instance
   in the appendix B, we have considered the case of a general
hypergeometric function whose
   argument is a matrix.]

The Itzykson-Zuber integral \cite{Itzykson} is an integral over
the unitary group $U(n)$,
\bq\label{3.2}
 \int_{U(n)} e^{{\rm tr} ( u a u^{\dag} b )} du = C_N{{\rm det }( e^{a_i
b_j})\over
 {\Delta(a) \Delta(b)}}
\eq
where $\Delta(a) = \prod_{i<j} (a_i - a_j)$, and similarly $\Delta(b)$, are
the Vandermonde determinants
of the eigenvalues of the Hermitian matrices $a$ and $b$.
The constant  is found to be $C_N = \prod_{j=1}^N (j - 1)!$.
By expressing ${\rm det}(e^{a_ib_j})$ as an alternating sum over
the symmetric group $S_n$, one may write
\bq\label{3.3}
\int_{U(n)} e^{<{\rm Ad}(u)\cdot a|b>} du = C_N{\sum \epsilon(w) e^{<w\cdot 
a|b>}
\over{\Delta(a)\Delta(b)}}
\eq
where ${\rm Ad}(u)\cdot a = u a u^{-1}$ is the adjoint action of $u\in
U(n)$ on the matrix $a$.
In this formula the matrix $a$ is diagonal,
$a = {\rm diag}(a_1,...,a_n)$, $w$ is a permutation and  $\epsilon(w)$ is
its signature.
We have used the notation $<a|b> = {\rm tr} (a b)$.

This formula is a special case of a more general formula due to Harish-Chandra
\cite{Harish-Chandra}. Let $G$ be a compact connected Lie group.
Then Harish-Chandra's result reads
\cite{Harish-Chandra,AltItzykson,GrossRichards1}
\bq\label{HC}
\int_G e^{<{\rm Ad}(g)\cdot a|b>} dg = {\sum_{w\in W} ({\rm det} w)
e^{<w\cdot a|b>}
\over{\Delta(a)\Delta(b)}}
\eq
where $a, b$, $h$ belong to a Lie algebra $h$, and for any $H\in h$,
\bq\label{3.5}
\Delta(H) = \prod_{\alpha\in \Delta_{+}} \alpha(H)
\eq
$\Delta_{+}$ is  the collection of the positive roots,
$w$ is the finite reflection group, called the Weyl (or Coxeter) group
and $h$ is the Cartan subalgebra.

The Harish-Chandra formula has been interpreted more
recently as an integration over an orbit, with a symplectic structure.
This structure implies that the saddle point method is exact, provided one
sums over all the critical points
 \cite{Duistermaat,Witten}.
This is also related to the localization theorems
\cite{Atiyah}.

As an illustration of the semi-classical nature of this formula,
we consider again the  integral over the unitary goup $U(N)$,
\bq\label{3.6}
I = \int {\rm d}g e^{{\rm tr}(a g b g^{\dag})}
\eq
Representing an element $g$ of $U(N)$ as
\bq\label{3.7}
g = g_0 e^{i X}
\eq
where $X$ is an element of the  Lie algebra, we expand the argument of the
exponential.
up to second order in $X$, and impose the stationarity condition,
\bq\label{3.8}
{\rm tr} [a g_0  i X b g_0^{\dag}] + {\rm tr}
[a g_0 b ( - i X) g_0^{\dag}]
= 0
\eq
for any $X$, i.e.
\bq\label{3.9}
[ b, g_0^{\dag} a g_0] = 0
\eq
If $a$ and $b$ are diagonal, it implies that $g_0$ should be a unitary
permutation matrix,
$g_0 = p$.
Then, the integral $I$ becomes
\bq\label{3.10}
I = \sum_p e^{{\rm tr}(a p b p^{-1})}
\int dX \exp[ {\rm tr}( -{1\over{2}}a p X^2 b p^{-1}
- {1\over{2}}a p b X^2 p^{-1} + a p X b X p )]
\eq
Performing the Gaussian integral over $X$ one recovers the Itzykson-Zuber
formula. (This is not a derivation of course, but a way of verifying that
for this integral
the one-loop approximation is exact
provided one sums over all the saddle-points).

A similar saddle point technique may be applied to the symplectic
case.
We take $g \in Sp(N)$ in (\ref{3.6}) and for $a$ and $b$
diagonal matrices with chiral eigenvalues, $a = {\rm diag}(a_1, ..., a_N,
-a_1, ..., -a_N)$
and similarly for $b$.
Then the same calculation yields
\bq\label{3.11}
I = C {{\rm det}[ 2 { \sinh }(2 a_i b_j)]\over{\prod_{1\le i < j \le N}
(a_i^2 - a_j^2) (b_i^2 - b_j^2)\prod_{1\le k \le N}(a_k b_k)}}
\eq
and this result will be repeatedly used below.

\section{DENSITY OF STATE}

Using the method developped for the unitary case
\cite{BH1,BHZ,BH2,BH3,Kazakov},
we consider now the density of states with an  external source matrix.
Indeed for a random
Gaussian $V$, the resulting probability for $H$ is a Gaussian with an
external matrix
source $A$ linearly coupled to $V$.
We assume that  the  matrix $A$ belongs to the
Lie algebra of the symplectic group and,
 without a loss of generality, we may take it as a diagonal matrix,
$A = {\rm diag}(a_1,...,a_N, -a_1,...,-a_N)$. We take this $A$ as the
unperturbed $H_0$.
The probability distribution of the random matrix $V$ is assumed to be
Gaussian.

The density of states $\rho(\lambda)$ is given by
\bq\label{4.1}
\rho(\lambda) = {1\over{N}}< \sum_{\alpha=1}^N \delta (\lambda -
\lambda_{\alpha}) >
\eq
The probability distribution $P(M)$ of the random matrix $M$ is
\begin{eqnarray}\label{4.2}
P(M) &=& \exp [ - {\rm tr} M^2 - {\rm tr} M A ]\nonumber\\
     &=& \exp [ - \sum \lambda_i^2  - {\rm tr}( g \Lambda g^{\dag} A )]
\end{eqnarray}
where $\Lambda = {\rm diag}(\lambda_1,...,\lambda_N,-\lambda_1,...,-\lambda_N)$
and $g$ is an element of the symplectic group.
(Note that we use here a normalization which differs from our
 previous treatment \cite{BH1,BH2,BH3} of the unitary case).
With the present normalization the edge of the density of states becomes
of order $\sqrt{N}$ ;
in the large N limit, the support of the density of states
lies in an interval of  order $[-\sqrt{2N},\sqrt{2N}]$.

The density of states is obtained as the  Fourier transform of the evolution
operator
\bq\label{4.4}
\rho(\lambda) = \int_{-\infty}^{\infty} {dt\over{2 \pi}}
e^{- i t \lambda} U_A(t)
\eq
and using the symplectic measure $\Delta^2(\lambda)$ with  $\Delta(\lambda)
= \prod (\lambda_i^2 - \lambda_j^2) \prod \lambda_k$ \cite{HM},
and  the previous
Harish-Chandra integral
formula (\ref{3.11}), we obtain
\begin{eqnarray}\label{4.3}
U_A(t) &=& {1\over{N}}\sum_{\alpha=1}^N
\int_{0}^{\infty} \prod_{i=1}^N{\rm d\lambda_i}
{\prod_{1\le i<j\le N}(\lambda_i^2 - \lambda_j^2) \prod_{1\le k \le N}
\lambda_k
\over{\prod_{1\le i<j\le N}(a_i^2 - a_j^2)\prod_{1\le k \le N}a_k}}
e^{- \sum \lambda_i^2 + i t \lambda_{\alpha}} \nonumber\\
&\times & {\rm det}[ {\rm sh} ( 2 \lambda_i a_j)]
\end{eqnarray}
(up to a  factor 2 in (\ref{3.11})
 absorbed in the coefficient $C$).

By the reflexion symmetry  $\lambda_i\rightarrow - \lambda_i$, and by the
exchange symmetry between $\lambda_i$ and $\lambda_j$, we
can extend  the integrations over the $\lambda_i$'s from $-\infty$ to $\infty$
. The  change of sign for $i t \lambda_i$ can be absorbed since
 we can change $t \rightarrow - t$ by parity. Then the expression for $U_A(t)$
 simplifies to
\bq\label{4.5}
 U_A(t) = {1\over{N}}
 \sum_{\alpha=1}^N \int_{-\infty}^{\infty} \prod_{i=1}^N {\rm d}\lambda_i
{\prod_{1\le i<j\le N}(\lambda_i^2 - \lambda_j^2) \prod_{1\le k \le N}
\lambda_k
\over{\prod_{1\le i<j\le N}(a_i^2 - a_j^2)\prod_{1\le k \le N}a_k}}
e^{- \sum \lambda_i^2 + i t \lambda_{\alpha} + 2 \sum_{i}^N a_i \lambda_i}
\eq

Before proceeding to the explicit integrations over the $\lambda_i$'s in
$U_A(t)$,
let us investigate the expression for the density of states $\rho(\lambda)$
, in the absence of any external source, which one may then calculate by
the usual method with orthogonal polynomials
.
The N-point level distribution is given
\bq\label{4.6}
\rho_N( \lambda_1,...,\lambda_N) = C \prod_{1\le i<j\le N}(\lambda_i^2 -
\lambda_j^2)^2
 \prod_{1\le k \le N} \lambda_k^2 \exp[ - \sum \lambda_i^2 ]
\eq
Using orthogonal polynomials method \cite {Mehta}, this distribution may be
written as
\bq\label{4.7}
\rho_N( \lambda_1,...,\lambda_N) = {\rm det} [ K_N(\lambda_i,\lambda_j)]
\eq
with the kernel $K_N(\lambda_i,\lambda_j)$ given by a sum of Laguerre
polynomials, and in
particular the density of states $\rho(\lambda)= K_N(\lambda,\lambda)$ is
equal to

\bq\label{4.8}
\rho(\lambda) = {1\over{N}} \sum_{n=0}^{N-1} {n!\over{\Gamma({3\over{2}} + n)}}
L_n^{({1\over{2}})}(\lambda^2)
L_n^{({1\over{2}})}(\lambda^2)\lambda^2 e^{-\lambda^2}.
\eq
 $L_n^{(1/2)}(x)$ are associated (generalized) Laguerre polynomials,
orthogonal with the normalization

\bq\label{4.9}
 \int_{0}^{\infty} e^{-x} x^{{1\over{2}}} L_n^{({1\over{2}})}(x)
 L_m^{({1\over{2}})}(x) dx = {\Gamma({3\over{2}} + n)\over{n!}} \delta_{n,m} ;
\eq
leading to  $ L_0^{(1/2)} (x) = 1, L_1^{(1/2)}(x) = 3/2 - x$, $
L_n^{(1/2)}(x) = \sum_{r=0}^n (-1)^r [\Gamma(n + 3/2)/
\Gamma(n - r + 1)\Gamma( r + 3/2)]x^r/r!$.

The asmptotic behavior of these Laguerre polynomials in the limit $N\rightarrow
\infty$ is \cite{Erdelyi}
\bq\label{4.10}
L_N^{({1\over{2}})}(x) = {1\over{\sqrt{\pi x}}}e^{{1\over{2}}x}
\sin (2 \sqrt{N x}) + O({1\over{\sqrt{N}}}).
\eq
With the Christoffel-Darboux identity, we have
\begin{eqnarray}\label{4.11}
& &\sum_{n=0}^{N-1} {n!\over{\Gamma({3\over{2}} + n)}}
L_n^{({1\over{2}})}(x) L_n^{({1\over{2}})}(y)\nonumber\\
&=& - {N!\over{\Gamma({1\over{2}} + N)}}
{L_{N}^{({1\over{2}})}(x) L_{N-1}^{({1\over{2}})}(y)
- L_{N-1}^{({1\over{2}})}(x) L_{N}^{({1\over{2}})}(y)\over{x - y}}
\end{eqnarray}
and using  the large $N$ aymptotic behavior
(\ref{4.10}), we obtain
\begin{eqnarray}\label{4.12}
&& x e^{-x}[\sum_{n=0}^{N-1}{n!\over{\Gamma({3\over{2}} + n)}}
 L_n^{({1\over{2}})}(x) L_n^{({1\over{2}})}(x)]\nonumber\\
 &=&  x e^{-x}[L_{N}^{({1\over{2}})}(x) {d\over{d x}}L_{N-1}^{({1\over{2}})}(x)
- L_{N-1}^{({1\over{2}})}(x) {d\over{d x}}L_{N}^{({1\over{2}})}(x)]\nonumber\\
&\simeq&{1\over{\pi}}[ {\sqrt{N}\over{\sqrt{x}}}\sin (\sqrt{{x\over{N}}}) -
{1\over{4\sqrt{N x}  }}
\sin(4\sqrt{N x})]
\nonumber\\
&\simeq&  {1\over{\pi}}[1 - {1\over{4\sqrt{N x} }}
\sin(4\sqrt{N x})]
\end{eqnarray}
Thus, putting $x = \lambda^2$, we get
\bq\label{4.13}
\rho(\lambda) =  {1\over{\pi}}[ 1 -
{\sin(4 \sqrt{N} \lambda)\over{4 \sqrt{N} \lambda}}].
\eq
This result is identical to  (\ref{2.13}),
when we substitute $\omega_0 = \frac{\pi}{4 N}$ and $E = \lambda/\sqrt{N}$.

We now return to  the integral (\ref{4.5}) over the $\lambda_i$'s in  the
presence of the
external source.
We first  replace  $t \rightarrow 2 t$ and following \cite{BH1,BHZ},
for fixed $\alpha$, we  denote by $b_i$  the sum $b_i = a_i + i t
\delta_{i,\alpha}$ . One then
uses the integral
\bq\label{4.14}
 \int_{-\infty}^{\infty} {\rm d}\lambda_i \prod_{i<j}(\lambda_i^2 -
\lambda_j^2) \prod_{i=1}^N
 \lambda_i e^{- \sum (\lambda_i - b_i)^2 }
 = \prod_{i<j} (b_i^2 - b_j^2)\prod_{i=1}^N b_i ;
 \eq
(after the translation  $\lambda_i \rightarrow\lambda_i+b_i$ the result
follows easily from antisymmetry
under permutation of the $b_i$'s, parity in  $b_i$, and counting the degree
of the resulting polynomial
in those $b$'s).

 Using the normalization  $U_A(0) = 1$, and writing the sum over
 $\alpha$ as a contour integral around the $a_j^2$'s in the complex
$u$-plane, we obtain
 \bq\label{4.15}
 U_A(t) = {1\over{N}}\oint {du\over{2 \pi i}}
 \prod_{j=1}^N
 {((\sqrt{u} + i t)^2 - a_j^2)\over{(u - a_j^2)}}
 {1\over{(\sqrt{u} + i t)^2 - u}}(1 + {i t\over{\sqrt{u}}})
 e^{- t^2 + 2 i t \sqrt{u}}.
 \eq

 Fourier transforming  $U_A(t)$ we obtain the density of states
$\rho(\lambda)$

 \bq\label{4.16}
 \rho(\lambda) = \int_{-\infty}^{\infty} U_A(t) e^{- 2 i t \lambda}
 {dt\over{2 \pi}}
 \eq
 Let us verify the consistency with the orthogonal polynomial result
(\ref{4.8}) when the external source vanishes.
Then
\bq
U_0(t) = {1\over{N}}
\oint {du\over{2 \pi i}} {(\sqrt{u} + i t)^{2N}\over{u^N}}
{1\over{(\sqrt{u} + i t)^2 - u}}
( 1 + {i t\over{\sqrt{u}}}) e^{- t^2 + 2 i t \sqrt{u}}
\eq
and
\begin{eqnarray}\label{4.18}
\rho(\lambda) &=&
\int_{-\infty}^{\infty} {dt\over{2 \pi}}
U_0(t) e^{- 2 i t\lambda}\nonumber\\
&=& {i (-1)^N\over{N}} \int_{-\infty}^{\infty}{dx\over{2 \pi}}
\oint{du\over{2 \pi i}} ({x\over{\sqrt{u}}})^{2N + 1} {1\over{u + x^2}}
e^{-x^2 - 2 i x \lambda - u + 2 \lambda \sqrt{u}}
\end{eqnarray}
(we have shifted $t$ to $ t = x + i \sqrt{u}$). The integration
contour is a small loop around the origin $u = 0$.
Expanding $1/(u + x^2) = x^{-2}( 1 - {u\over{x^2}} + \cdots )$,
and noting that
\bq\label{4.19}
\oint {du\over{2 \pi i}} {1\over{u^{n + 1} \sqrt{u}}} e^{- u + 2
\lambda \sqrt{u} } = { 2 \Gamma({3\over{2}})\over{\Gamma({3\over{2}} + n)}}
(-1)^n \lambda L_n^{({1\over{2}})} (\lambda^2)
\eq
and
\bq\label{4.20}
\int_{-\infty}^{\infty} {dx\over{2 \pi}} x^{2n + 1} e^{- x^2 - 2 i x \lambda}
= - {i n!\lambda \over{2 \sqrt{\pi}}}L_n^{({1\over{2}})}(\lambda^2)
e^{-\lambda^2}
\eq
we v erify the consistency of the integral representation  (\ref{4.18})
with the orthogonal polynomial result (\ref{4.8}). The integral
representation  (\ref{4.20}),
gives an easy way to recover the large $n$ asymptotic behavior (\ref{4.11}) of
the Laguerre polynomials through the steepest descent method.
Note that the chiral invariance leads as expected to an even density of states.

In the presence of the external source, the density of states becomes
\bq
\rho(\lambda) =
{i (-1)^{N - 1}\over{N}} \int_{-\infty}^{\infty}{dx\over{2 \pi}}
\oint{du\over{2 \pi i}} ({x\over{\sqrt{u}}})
\prod_{\gamma=1}^N ({x^2 + a_{\gamma}^2 \over{u - a_{\gamma}^2}})
 {1\over{u + x^2}}
e^{-x^2 - 2 i x \lambda - u + 2 \lambda \sqrt{u}}\nonumber\\
\eq
(We shall analyze this density of states as the limit of $\lambda \rightarrow
\mu $ of the kernel $K_N(\lambda,\mu )$ in the next section).
We note here that if we make the change of variable, $u = v^2$,
we have
\bq
\rho(\lambda) = {i (-1)^{N -1}\over{N}}
\int_{-\infty}^{\infty} {dx\over{2 \pi}}
\oint{dv\over{2 \pi i}}
\prod_{\gamma=1}^N ({x^2 + a_{\gamma}^2 \over{v^2 - a_{\gamma}^2}})
 {x\over{v^2 + x^2}}
e^{-x^2 - 2 i x \lambda - v^2 + 2 \lambda v}\nonumber\\
\eq
where the contour encloses all the $\pm a_{\gamma}$.
For example when $N= 1$, we obtain
\begin{eqnarray}
\rho(\lambda) &=& {\lambda\over{2 \sqrt{\pi}}}{{\rm sinh}( 2 \lambda a_1 )
\over{a_1}}e^{- a_1^2 - \lambda^2}\nonumber\\
&=& {\lambda\over{4 \sqrt{\pi} a_1}}[ e^{-(\lambda - a_1)^2} - e^{-(\lambda
+ a_1)^2}]
\end{eqnarray}

Near the origin $\lambda = 0$, the density of states $\rho(\lambda)$
becomes $\rho(\lambda)
\sim {\lambda^2\over{\pi}} {\rm exp}( - a_1^2 - \lambda^2)$.
Thus the density of states is an even function and it vanishes near the origin
as $\lambda^2$.

%*********************
\vskip 5mm
\section{CORRELATON FUNCTION}
\vskip 5mm

 The $n$-level correlation functions are expressed as the determinant of
the kernel
 $K_N(\lambda,\mu )$, exactly as for Hermitian random matrices
 in a source. Indeed the formula of (\ref{4.14}) is quite similar to
 (3.8) of \cite{BH3}. Therefore, all the derivations of the
 Hermitian matrix model in \cite{BH1,BH2,BH3}, may be repeated almost
identically for the $Sp(N)$
 case.

 For instance, the two-level correlation function $\rho^{(2)}(\lambda,\mu )$
is obtained as
 the double Fourier transform of $U_A(t_1,t_2)$,
 \bq\label{5.1}
 \rho^{(2)}(\lambda,\mu ) = \int\int {dt_1 dt_2\over{(2 \pi )^2}}
 e^{- i t_1 \lambda - i t_2 \mu } U_A(t_1,t_2)
 \eq
in which
 $U_A(t_1,t_2)$ is defined as
 \bq\label{5.2}
 U_A(t_1,t_2) = < {1\over{N}} \sum_{\alpha=1}^N e^{i t_1\lambda_{\alpha}}
 {1\over{N}} \sum_{\beta=1} e^{i t_2 \lambda_{\beta}} >
 \eq
 Using the integral formulae (\ref{3.10}) and (\ref{4.14}), we find
 \begin{eqnarray}\label{5.3}
 U_A(t_1,t_2) &=& {1\over{N^2}}\sum_{\alpha\neq \beta}
 {(a_{\alpha} + i t_1)^2 - ( a_{\beta} + i t_2)^2\over{
 (a_{\alpha}^2 - a_{\beta}^2)}}
 \prod_{\gamma \neq (\alpha,\beta)}
 {(a_{\alpha} + i t_1)^2 - a_{\gamma}^2\over{
 a_{\alpha}^2 - a_{\gamma}^2 }}
 \nonumber\\
 &\times &\prod_{\gamma \neq (\alpha,\beta)}
 {(a_{\beta} + i t_1)^2 - a_{\gamma}^2\over{
 a_{\beta}^2 - a_{\gamma}^2 }}
 (1 + {it\over{a_{\alpha}}})(1 + {i t_2\over{a_{\beta}}})
 e^{- t_1^2 - t_2^2 + 2 i \sqrt{u} t_1 + 2 i \sqrt{v} t_2}
 \nonumber\\
 \end{eqnarray}
 (wa have subtracted the term  $\alpha = \beta$ in $U_A(t_1,t_2)$).
 The summation over $\alpha$ and $\beta$
 is expressed by contour integration
over two complex variables $u$ and $v$,
 \begin{eqnarray}\label{5.4}
 U_A(t_1,t_2) &=& {1\over{N^2}} \oint {du dv\over{(2 \pi i)^2}}
 \prod_{\gamma=1}^N
 {(\sqrt{u} + i t_1)^2 -  a_{\gamma}^2\over{
 (u - a_{\gamma}^2)}}
 \prod_{\gamma = 1}^N
 {(\sqrt{v} + i t_2)^2 - a_{\gamma}^2\over{
 v  - a_{\gamma}^2 }}\nonumber\\
 &\times &
 {1\over{(\sqrt{u} + i t_1)^2 - v}}
 {1\over{(\sqrt{v} + i t_2)^2 - u}}
 (1 + {i t_1\over{\sqrt{u}}})(1 + {i t_2\over{\sqrt{v}}})
 \nonumber\\
 &\times &
 {(u - v) [(\sqrt{u} + i t_1)^2 - (\sqrt{v} + i t_2)^2]
 \over{[(\sqrt{u} + i t_1)^2 - u][(\sqrt{v} + i t_2)^2 - v]}}
 e^{- t_1^2 - t_2^2 + 2 i \sqrt{u} t_1 + 2 i \sqrt{v} t_2}
 \nonumber\\
 \end{eqnarray}

 We now shift $t_1\rightarrow t_1 + i \sqrt{u}$ and
 $ t_2\rightarrow t_2 + i \sqrt{v}$,
 note that
 \bq
 {(u - v) [ (i t_1)^2 - (i t_2)^2]\over{
 [(it_1)^2 - v][(it_2)^2 - u]}}
 = {[(it_1)^2 - u][(it_2)^2 - v]\over{[
 (it_1)^2 - v][(it_2)^2 - u]}} - 1
 \eq
 and divide $U_A(t_1,t_2)$ into two terms.  The first one, which comes from
(-1)
 gives the product of the two density of states $\rho(\lambda)\rho(\mu )$,
 and
 the second one  is a product of two integrals over $t_1,u$ and $t_2,v$
respectively.
 Therefore, we end up with
 \bq
 \rho^{(2)}(\lambda,\mu ) = K_N(\lambda,\lambda) K_N(\mu ,\mu )
 - K_N(\lambda,\mu )K_N(\mu ,\lambda)
 \eq
 with
 \begin{eqnarray}\label{5.7}
 K_N(\lambda,\mu ) &=& {1\over{N}}
 \oint {du\over{2 \pi i}}\int_{-\infty}^{\infty}
 {dt\over{2\pi}} e^{- t^2 - 2 i t \mu - u + 2 \sqrt{u} \lambda}
 \prod_{\gamma=1}^N {(i t)^2 - a_{\gamma}^2\over{u - a_{\gamma}^2}}{1\over{
 (it)^2 - u}}
 \nonumber\\
 &\times & ({it\over{\sqrt{u}}}) ;
 \end{eqnarray}
 (the contour in the $u$-plane encircles all the $a_{\gamma}^2$) .
 From this expression, we recover the previous result
(\ref{4.15},\ref{4.16}) for
 the density of states   as $\rho(\lambda)=
 K_N(\lambda,\lambda)$.

 The n-point correlation can be analyzed as in \cite{BH3} and
 leads here also to a determinant form,
 \bq
 \rho(\lambda_1,\cdots,\lambda_n) = {\rm det}[K_N(\lambda_i,\lambda_j)]
 \eq
 for an arbitrary external source.

 We now discuss Dyson's short-distance universality, within our model, i.e.
when we
vary the external source $A$,  through the explicit expression of the
 kernel $K_N(\lambda,\mu )$.

 We return first to the sourceless case,$a_{\gamma} = 0$, and
consider  the large N limit of $K_N(\lambda,\mu )$
 for $\lambda,\mu $ are order of one.
  We scale $t \rightarrow \sqrt{N} t$,
 and $u \rightarrow N u$. The kernel is then
 \bq\label{5.9}
 K_N(\lambda,\mu ) = - {1\over{\sqrt{N}}}\oint {du\over{2 \pi i}}
 \int {dt\over{2 \pi}}
 e^{- N (f(t) + g(u))} ({it\over{\sqrt{u}}}){1\over{t^2 + u}}
 \eq
 where $f(t) = t^2 + 2i\mu /\sqrt{N} - 2 \ln t$ and
 $g(u) = u - 2 \lambda \sqrt{u}/\sqrt{N} + \ln u$.
 In the large N-limit we have the saddle points, $t_c^{(1)} = 1 -
i\mu /2\sqrt{N}$,
 $t_c^{(2)} = - 1 - i\mu /2\sqrt{N}$,$u_c^{(1)} = -1 + i\lambda/2\sqrt{N}$
 and $u_c^{(2)} = -1 - i \mu /2\sqrt{N}$.
 Adding these saddle-points contributions, we obtain
 \bq\label{5.10}
 K_N(\lambda,\mu ) = {1\over{\pi}}[{\sin(2\sqrt{N}(\lambda - \mu ))
 \over{2 \sqrt{N}(\lambda - \mu )}} -
{\sin(2\sqrt{N}(\lambda + \mu ))
 \over{2 \sqrt{N}(\lambda + \mu )}}]
 \eq
 When $\lambda\rightarrow \mu $, we recover the expression (\ref{4.13}) of
the density of states.

  In the presence of the external source, we find a similar expression to
Eq.(\ref{5.9}),
  with
  \bq\label{5.11}
  f(t) = t^2 + {2it \mu \over{\sqrt{N}}} - {1\over{N}}\sum \ln (t^2 +
{a_{\gamma}^2
  \over{N}})
  \eq
  \bq\label{5.12}
  g(u) = u - {2 \lambda \sqrt{u}\over{\sqrt{N}}}
    + {1\over{N}} \sum \ln ( u - {a_{\gamma}^2\over{N}})
  \eq
 Using the definition , $f^{\prime}(t_c)=0,g^{\prime}(u_c) = 0$, of the
saddle-points
  we find
  \bq\label{5.13}
   K_N(\lambda,\mu ) = {1\over{\pi}}[{\sin(2 t_{c0}\sqrt{N}(\lambda - \mu ))
 \over{2 \sqrt{N}(\lambda - \mu )}} -
{\sin(2 t_{c0}\sqrt{N}(\lambda + \mu ))
 \over{2 \sqrt{N}(\lambda + \mu )}}]
 \eq
 where $t_{c0}$ is a solution of
 \bq\label{5.14}
   {1\over{N}} \sum_{\gamma = 1}^N {1\over{t_{c0}^2 + {a_{\gamma}^2\over{N}}}}
  = 1
  \eq
 Here we have assumed that the order of $a_{\gamma}$ is
  $a_{\gamma}^2 \leq O(N)$.
  The derivation of (\ref{5.13}) is the same as
  (\ref{5.10}).
  The difference is  just a normalization, due to the change of the
  saddle-point $t_{c0}$.
  Note that the support of the
  density of states  is inside the
  interval of $-\sqrt{N}$ and $\sqrt{N}$. Therefore,
   universality holds provided that the eigenvalues of the external source
matrix
   are located in an interval of the
  same order of magnitude ; ( it has to be of same order as the support of
the density of states for the zero external source
  case). In this universal regime for the correlation function, the
energies $\lambda$ and $\mu $ are
  assumed of order  one, namely $\lambda/\sqrt{N} \rightarrow 0$
  in the large N limit. Since $\lambda$ can be order of $\sqrt{N}$,
  the universal behavior of (\ref{5.13}) appears only near the origin,
 as exepcted, and is differnt from the non-universal bulk behaviour.
  The universality found here, is similar to that found in other chiral
random matrix models
 studied in \cite{BHZ}; there  it was the density of states which took a
universal
 form near the origin, and became independent of the external source or of the
 non-Gaussian distributions (as seen by a rescaling of the energy).

  When the  $a_{\gamma}^2$ spread over an interval larger than $N$,
  the universal form of (\ref{5.13}) does not hold any more, since the
  saddle-point method of (\ref{5.14}) breaks down.
  In our formulation, the amplitude of the random matrix is fixed.
  We now change the strength of the external source. For this purpose,
  we introduce a parameter $C$, which determines the strength
of the external source $H_0$, respective  to the random
  potential $V$. Let us  consider the example of the external source,
  \bq\label{5.15}
 a_{\gamma} = C {\rm tan} [{(2\gamma - 1)\pi\over{2 ( 2 N + 1)}}]
 \eq
 where $\gamma = 1, 2,..., N$, and $C$ is a parameter.
 When $\gamma/( 2N + 1) << 1$, we have
 \bq\label{5.16}
 a_{\gamma} = {C \pi \over{ ( 2 N + 1)}} ( \gamma - {1\over{2}})
 \eq
 and the eigenvalues of the external source are equally distributed,  and
thus reproduce
 the excitation spectrum inside the clean superconductor vortex, if we identify
  $\hbar \omega_0 = C\pi/(2N + 1)$ in (\ref{2.3}).
 Since we are interested in
 the behavior near the origin for the energies  $\lambda$ and $\mu $ in
$K_N(\lambda,
 \mu )$, this linear approximation of the ${\rm tan}(x) \simeq x$ is
 valid.

 More generally, using the   formula,
 \bq\label{5.17}
 \prod_{m = 1}^N ( x^2 + {\rm tan}^2 [{( 2 m - 1) \pi\over{2(2 N + 1)}}])
 = {1\over{2( 2 N + 1) }} [ ( 1 + x)^{2 N + 1} + ( 1 - x)^{2 N + 1}]
 \eq
 we may replace  the product involving the  $a_\gamma$ in (\ref{5.7}),
 and  obtain
 \begin{eqnarray}\label{5.18}
 K_N(\lambda,\mu ) &=& - {1\over{N}} \oint {du\over{2 \pi i}}
 \int {dt\over{2 \pi}}
 \left[ {( 1 + { t\over{C}})^{2N + 1} +
  ( 1 - { t\over{C}})^{2N + 1}\over{
  ( 1 + {i \sqrt{ u} \over{C}})^{2N + 1} +
  ( 1 - {i \sqrt{u} \over{C}})^{2N + 1}}}\right ]
   \nonumber\\
  &\times & {1\over{t^2 + u}} ({i t\over{\sqrt{u}}})
 e^{-  t^2 - 2 i  t \mu -  u + 2 \sqrt{u } \lambda}
 \end{eqnarray}

 This expression is exact for finite $N$ when the $a_{\gamma}$ are given by
 (\ref{5.15}).
 It is easily seen that when $C\rightarrow 0$, we recover the previous
external source free result (\ref{5.9}).

 For non-zero $C$ we have two different cases according to the size  of $C$,\\
 i) if $C$ is of order  $\sqrt{N}$, we recover the universal behavior of
(\ref{5.13}). (We simply scale
 $t\rightarrow \sqrt{N}t$, and $u \rightarrow Nu$,
 and use the saddle point method). We recover then the result
  (\ref{5.13}), with
 \bq
 t_{c0} = {1\over{2}}( \sqrt{{C^2\over{N}} + 4} - {C \over{\sqrt{N}}} )
 \eq.\\
  ii) in the second case $C \sim O(N)$,
 we change  $u = v^2$ in (\ref{5.18}) :
 \begin{eqnarray}\label{5.20}
 K_N(\lambda,\mu ) &=& - {1\over{N}} \oint {dv\over{2 \pi i}}
 \int {dt\over{2 \pi}}
 \left[ {( 1 + { t\over{C}})^{2N + 1} +
  ( 1 - { t\over{C}})^{2N + 1}\over{
  ( 1 + {i v \over{C}})^{2N + 1} +
  ( 1 - {i v \over{C}})^{2N + 1}}}\right ]
   \nonumber\\
  &\times & {i t\over{t^2 + v}}
 e^{-  t^2 - 2 i  t \mu -  v^2 + 2 v \lambda}
 \end{eqnarray}

 In the large N-limit (\ref{5.20}) reduces to
 \begin{eqnarray}\label{5.21}
  K_N(\lambda,\mu ) &=& - {1\over{N}}
  \oint {dv\over{2 \pi i}}\int {dt\over{2 \pi}}
  {{\rm cosh} ( {2 N t\over{C}} )\over{{\rm cos} ( {2 N v\over{C}}
  )}}{i t\over{t^2 + v^2}}
  \nonumber\\
  &\times & e^{- t^2 - 2 i  t \mu - v^2 + 2 v \lambda}
  \end{eqnarray}
  If we rescale $t \rightarrow \sqrt{N} t$ and $v \rightarrow \sqrt{N} v$
    poles appears in the $v$-plane at
  $ v = C ( m - {1\over{2}})\pi/ 2 N^{3/2}$
  for which  ${\rm cos}( 2 N^{3/2} v/C) = 0$ ;  $m$ is an integer.

 Within  the residues of these poles, we have the Gaussian factors
  \bq
  e^{- N (v - {\lambda\over{\sqrt{N}}})^2} =
  e^{-({C( m - {1\over{2}})\pi\over{2 N}} - \lambda )^2}
  \eq

  There are thus two different regions for $\lambda$. We first consider
  $\lambda \sim O(\sqrt{N})$, i.e. the bulk case.
  Then $\lambda/\sqrt{N} >> \sqrt{N}/C$, and the saddle
  point of $t$ is $t_0 = - i \lambda/\sqrt{N}$. We set $\lambda = \mu $, and
for this
 this saddle point, we have the density of states $\rho(\lambda) =
  K_N(\lambda,\lambda)$
 \begin{eqnarray}\label{5.23}
 \rho(\lambda) &=& - {1\over{2\sqrt{\pi N}}}
 \sum_m (-1)^m
 [{\lambda{\rm cos} ( {2 N\lambda\over{C}})
 \over{- \lambda^2 + ({C (m - {1\over{2}})\pi\over{2 N}})^2}}]
 \nonumber\\
 &\times & e^{-({C( m - {1\over{2}})\pi\over{2 N}}- \lambda )^2}
 \end{eqnarray}

 Note that the zeros of the denominator are cancelled by zeros of
 the numerator ${\rm cos}( 2 N \lambda/ C)$.
 When $C/N >> 1$,
  the exponential factor of (\ref{5.23}) damps the result and we may
 set $\lambda = C( m - 1/2)\pi/(2 N)$,
 then
 \bq
  {\lambda {\rm cos} ( {2 N\lambda\over{C}})
 \over{- \lambda^2 + ({C (m - {1\over{2}})\pi\over{2 N}})^2}}
 = - (- 1)^m {N\over{C}}
 \eq
 Thus we find
 \bq
 \rho(\lambda) = {\sqrt{N}\over{2 C \sqrt{\pi}}}
 \sum_m e^{-({C( m - {1\over{2}})\pi\over{2 N}}- \lambda )^2}
 \eq
 which is a sum of Gaussians :
 the density of states is just sum of Gaussian peaks around the levels
  $\lambda = C( m - 1/2)\pi/2 N$. This result is valid provided
 $\lambda$ is not too close to the origin.

 When $\lambda$ is near the origin, namely $\lambda \sim O(1)$, we have
 $\lambda/\sqrt{N} \sim \sqrt{N}/C$ since $C$ is order $N$.
 In this case, we have to take into account the term ${\rm cosh}( 2N t/C)$
 in the saddle-point equation. The saddle-point becomes
 $t_0 = - i \lambda/\sqrt{N} \pm \sqrt{N}/C$.
 Using this value in (\ref{5.21}), we find
 \bq
 \rho(\lambda) = {\rm Re} \sum (-1)^n [ {\lambda + i {N\over{C}}\over{
 -(\lambda + i {N\over{C}})^2 + ({C\over{2 N}}( n - {1\over{2}})\pi )^2}}]
 e^{-[{C\over{2 N}}( n - {1\over{2}})\pi - \lambda ]^2}
 \eq
 Near $\lambda = 0$, the density of states behaves as $\rho(\lambda) \sim
 \lambda^2$.
 For example, when $C = N$, we have
 $\rho(\lambda ) \sim \pi \lambda^2 {\rm exp}
 ( - \pi^2/16)/(1 + \pi^2/16)$, and $\rho(\lambda)$
 has  peaks at $\lambda = \pm \pi/4, \pm 3 \pi/4,...$.
 Thus, when $C \sim N$, which means that the random potential is weak
 compared to the external source,
 the density of states takes a form which is different from the universal
 form (\ref{5.13}).

\vskip 5mm

\section{CROSSOVER FROM THE CLEAN LIMIT TO THE
DIRTY LIMIT}

In the clean case, when $k_F r_i \sim k_F \xi >> 1$, we approximate $A_{nm}^i$
by (\ref{2.19}). Then, the periodic spectrum is obtained
\bq\label{6.1}
 E_n = - \hbar \omega_0 ( n - {1\over{2}} + \tilde z (- 1)^n )
\eq
When the number of  impurities inside the
vortex $N_i \simeq 1$, we are in the superclean case, and
 $\tilde z$ in (\ref{6.1}) is a function of the position of this
impurity.
If $1 < N_i < N_{ic}$, we are in a clean case, and we need to average over
the different values of $\tilde z$.

In the previous section, we have
discussed the case where the deterministic term $E_n = - \hbar \omega
( n - {1\over{2}})$ is coupled to the random matrix $A$.
Here we consider for the external source the matrix whose eigenvalues are
(\ref{6.1}).

We may then use the
 formula
\bq\label{6.2}
\prod_{r = 1}^N [x^2 + {\rm tan}^2 ( {r \pi\over{2 ( N + 1)}})]
= {1\over{4 (N + 1) x}} [ (1 + x )^{2(N + 1)} - ( 1 - x)^{2(N + 1)}]
,\eq
 which is similar to (\ref{5.17}). (
We assume that $N$ is odd  here).
We divide this product into two parts, r-odd and r-even .
The r-even part is obtained immediately from (\ref{6.2}) as
\bq\label{6.3}
\prod_{r = even}^{N - 1} [x^2 + {\rm tan }^2 ( {r \pi\over{2 ( N + 1)}} )]
= {1\over{2( N + 1) x}} [ (1 + x)^{N + 1} - (1 - x)^{N + 1}]
\eq
Dividing (\ref{6.2}) by (\ref{6.3}), we obtain the expression for the r-odd
part,
\bq\label{6.4}
\prod_{r = odd}^N [x^2 + {\rm tan}^2 ( {r \pi\over{2 ( N + 1)}})]
= {1\over{2}}[ (1 + x)^{N + 1} + ( 1 - x)^{N + 1} ]
\eq
We now consider the eigenvalues  (\ref{6.1}) as an external source,

\bq\label{6.5}
a_{\gamma} = [C \pi/2(N + 1)](\gamma -
{1\over{2}} + \tilde z ( - 1)^{\gamma}),
\eq
 and introduce $z_0 = C \pi/4(N + 1)$, $z = - C \pi \tilde z/2(N + 1)$.
Using the expressions
 (\ref{6.3}) and (\ref{6.4}), we write
\begin{eqnarray}\label{6.6}
&&({x - iz_0 - iz\over{C}}) \prod_{r = even}^{N - 1}
[({x - iz_0 - iz\over{C}})^2 + {\rm tan}^2({r \pi\over{2 ( N +
1)}})]\nonumber\\
&\times &
\prod_{r = odd}^N [ ({x - z_0 + z\over{C}})^2 + {\rm tan}^2 ({r \pi\over{2(
N + 1)}})]\nonumber\\
&=& {1\over{4( N + 1)}}
[ ( 1 + {x - iz_0 - iz\over{C}})^{N + 1}
- ( 1 - {x - iz_0 - iz\over{C}})^{N + 1}]
\nonumber\\
&\times &
[( 1 + {x - iz_0 + iz\over{C}})^{N + 1} + ( 1 + {x - iz_0 + iz\over{C}})^{N
+ 1}]
\end{eqnarray}
In the large N limit, when $r\pi/(2(N + 1)) << 1$,  the
left hand side of (\ref{6.5}) vanishes at $ x = - i(\pm r - 1/2 + z(-1)^r)
C \pi /[2(N + 1)]$ .
Noting that
\begin{eqnarray}\label{6.7}
\prod_{\gamma} (t^2 + a_{\gamma}^2) &=&
\prod_{\gamma} [ t - i(z_0 + z) + i {C\gamma \pi\over{2(N + 1)}}]
[ t - i(z_0 + z) - i {C\gamma \pi\over{2(N + 1)}}]\nonumber\\
&=& \prod_{\gamma} [ ( t - i(z_0 + z))^2 + C^2 {\rm tan}^2 ({\gamma \pi
\over{2(N + 1)}})]
\end{eqnarray}
we have
from (\ref{5.7}) (and the change of variable $\sqrt{u} = v$),
\begin{eqnarray}\label{6.8}
K_N(\lambda,\mu ) &=&
- {1\over{N}} \oint{dv\over{2 \pi i}}\int {dt\over{2 \pi}}
e^{-t^2 - 2 i t \mu - v^2 + 2 v \lambda} {it\over{t^2 + v^2}}
\nonumber\\
&\times &
{[( 1 + {i\over{C}}( - it -  z_0 -  z))^{N + 1}  -
( 1 - {i\over{C}}( - i t -  z_0 -  z))^{N + 1}]
\over{[( 1 + {i\over{C}}( v - z_0 - z))^{N + 1} -
( 1 - {i\over{C}}( v - z_0 - z))^{N + 1}]
}}\nonumber\\
&\times &{
[( 1 + {i\over{C}}( - i t -  z_0 +  z))^{N + 1}  +
( 1 - {i\over{C}}( - i t -  z_0 +  z))^{N + 1}]\over{
[( 1 + {i\over{C}}( v - z_0 + z))^{N + 1} +
( 1 - {i\over{C}}( v - z_0 + z))^{N + 1}]}}
\nonumber\\
\end{eqnarray}
As in the previous section, we have to distinguish two different cases, i)
$C \sim O(N)$
and ii) $C \sim O(\sqrt{N})$. There are also two regions $\lambda \sim
O(\sqrt{N})$ and
$\lambda \sim O(1)$.

 When $C \sim O(N)$, we exponentiate the factor in the bracket of (\ref{6.7}),
\begin{eqnarray}\label{6.9}
&&{{\rm sin}[ {N\over{C}}( - i t - z_0 - z) ]{\rm cos} [{N\over{C}}
( - i t - z_0 + z)]\over{
{\rm sin}[ {N\over{C}}( v - z_0 - z)] {\rm cos} [ {N\over{C}}
(v - z_0 + z)]}}\nonumber\\
&=& {{\rm sin} [ {2 N\over{C}} ( - i t - z_0)] - {\rm sin}
({2 N \over{C}} z)\over{
{\rm sin} [ {2 N\over{C}} ( v - z_0)] - {\rm sin}
({2 N \over{C}} z)}}
\end{eqnarray}
Since $z_0 = C \pi/[4(N + 1)]$, we have
\begin{eqnarray}\label{6.10}
  K_N(\lambda,\mu ) &=& - {1\over{N}}
  \oint {dv\over{2 \pi i}}\int {dt\over{2 \pi}}
  {{\rm cosh} ( {2 N t\over{C}} ) + {\rm sin}({2N\over{C}} z)
  \over{{\rm cos} ( {2 N v\over{C}}
  ) + {\rm sin}({2N\over{C}} z) }}{i t\over{t^2 + v^2}}
  \nonumber\\
  &\times & e^{- t^2 - 2 i  t \mu - v^2 + 2 v \lambda}
  \end{eqnarray}
When $z=0$, we obtain the same expression as in the previous
section  (\ref{5.21}).

Poles in the integral over $v$ are present at
$ v = z_0 + z +    C  n\pi/ N $ and $v = z_0  - z + (2 n + 1)
(C \pi/2 N)$ from (\ref{6.8}).
The residues $R$ for the first poles are
\bq\label{6.11}
R = (- 1)^n i t
 {{\rm sin}[ {N\over{C}}( - i t - z_0 - z)] {\rm cos} [ {N\over{C}}
( - i t - z_0 + z)]\over{
[ t^2 + (z_0 + z+ {C n \pi\over{N}})^2] {\rm cos} [ {2N\over{C}}
( z + {C n \pi\over{N}})]}} \nonumber\\
\eq
 When $\lambda = \mu \sim O(\sqrt{N})$, the saddle point becomes
$t_0 = - i \lambda/\sqrt{N}$ and we set this value in $R$.
The second pole gives a similar expression.
Furthermore, if $C/N >> 1$, we can approximate $\lambda \sim
z_0 + z +    C  n\pi/ N$. Then, we find that the residue $R$ in (\ref{6.10})
becomes $R = N/2C$.
Therefore, we have the density of states for $C/N >>1$,
and $\lambda \sim O(\sqrt{N})$,
\bq\label{6.12}
\rho(\lambda) = {\sqrt{N}\over{4 C \sqrt{\pi}}}
\sum_n [ e^{-( z_0 + z + {C n \pi\over{N}} - \lambda)^2}
+ e^{- ( z_0 - z + {C ( 2 n + 1)\pi\over{2 N}} - \lambda )^2}]
\eq
which is a sum of Gaussian distributions.

In the clean case we have several impurities and one may average this
density of states over $z$ .
As discussed in \cite{KL1}, the measure for this $z$-average is
\bq\label{6.13}
\rho_0(z) = A{\rm cos}^2[ { 2 N z \over{C}}]
\eq
where $A = 4N/C \pi$.
The matrix elements are represented by a quaternion, and the measure is
isomorphic
to the uniform measure over the four-dimensional sphere $s^{3}$, $Sp(1)
\sim S^3$.
In spherical coordinate this leads to
(\ref{6.13}).
When $C/N >> 1$, the integration of (\ref{6.12})
over $z$ with the measure (\ref{6.14})
yields
\begin{eqnarray}\label{6.14}
 <\rho(z)> &=&
 \int_{-\infty}^{\infty} e^{-(z_0 + z + {C n \pi\over{N}} - \lambda)^2}
 A {\rm cos}^2 ({2 N z\over{C}}) dz \nonumber\\
 &=& A {\sqrt{\pi}\over{2}} [ 1 - e^{- {4 N^2\over{C^2}}} {\rm cos}
 ( {4 N\over{C}} \lambda )]
 \end{eqnarray}
(in which  we have used $N/C << 1$).
 When $C/N \rightarrow \infty$, the exponential factor
 ${\rm exp}( - 4 N^2/C^2)$ becomes one and the density of states becomes ${\rm
 sin}^2( 2 N \lambda/C)$.
 This is indeed the result for the clean case found in \cite{KL1}. When $C/N$
 is finite, at the energy $\lambda = n \pi C/2N = n \hbar \omega_0$,
 the density of state becomes non-vanishing, except at the origin.
  At the origin $\lambda = 0$, we have a zero
 due to the factor $ i t = \lambda $ in (\ref{6.11}).
 It may be interesting to note this deviation from zero has been
 observed in the numerical work of \cite{KL1}.

 When $C$ becomes order of $\sqrt{N}$, we may apply the saddle point method
 in (\ref{6.8}). We take the saddle point equation (\ref{5.14}) for
 $a_{\gamma} = {\pi C\over{2 N}} (n - {1\over{2}} + ( -1)^{\gamma}
 \tilde z)$. However, the second and third term  $( -1/2 + (-1)^{\gamma}
 \tilde z)$ can be neglected since they remain of order one, compared to $n$
 which can be order of $N$. Thus we may approximate $a-{\gamma}$  by
 $a_{\gamma} = \pi^2 C^2 n^2/4 N^3$, which is independent of $z$.
 Therefore the average over $z$ does not change the result for the
 density of states, and we have a universal result for the kernel
 and for the density of states (\ref{5.13}).

 It is possible to take the average (\ref{6.10}) by the formula,
 \bq\label{6.15}
 \int_{-{\pi\over{2}}}^{{\pi\over{2}}} dz
 {a + {\rm sin} z\over{b + {\rm sin} z}} {\rm cos}^2 z
 = \pi [ {1\over{2}} + ( a - b) ( b - \sqrt{b^2 - 1})]
 \eq
 Then, changing the contour integration in the $v$-plane to the integration
 near the saddle-points  for $v$ in the large N limit, we have
 \begin{eqnarray}\label{6.16}
 K_N(\lambda,\mu ) &=& - {CA\pi\over{4 N^2}}
 \int{dt\over{2 \pi}}\int {dv\over{2 \pi}}({t\over{t^2 + v^2}})
 [ ({\rm cosh}({2 N t\over{C}}) - {\rm cos}({2 N v\over{C}})
 ) \nonumber\\
 &\times &  {\rm sin} ( {2 N v\over{C}})]
  e^{- t^2 - 2 i t \mu - v^2 + 2 v \lambda}
 \end{eqnarray}
 The term ${\rm sin}( 2N v/C )$ comes from the sum of  integration paths
 in opposite directions.
 We assume here that $\lambda$ and $\mu $ are of order  one, and
 $O(\lambda) \sim O({N\over{C}})$. After the shift $t \rightarrow
 \sqrt{N} t$, the saddle-point is at $t_0 = - i \mu /\sqrt{N}
 \pm \sqrt{N}/C$ and $v = \lambda/\sqrt{N} \pm i \sqrt{N}/C$ for the
 terms in (\ref{6.16}).
 Inserting the values of these saddle-points, we have
 \begin{eqnarray}\label{6.17}
 \rho(\lambda) &=& {N\over{C}}[ 1 - {C\over{4 N \lambda}} {\rm sin}
 ({4 N \lambda\over{C}})] + {C\over{2 N}}{\rm sin}^2 ({2 N \lambda\over{C}})
 \nonumber\\
 &-& {C \lambda^2\over{4 N ( \lambda^2 + {N^2\over{C^2}})}} [ 1 +
  e^{- {4 N^2\over{C^2}}}{\rm cos} ({4 N \lambda\over{C}})]
  \nonumber\\
  &+& {\lambda\over{4 (\lambda^2 + {N^2\over{C^2}})}}
  e^{- {4 N^2\over{C^2}}}{\rm sin} ( {4 N \lambda\over{C}})
  \end{eqnarray}

 The above expression reduces to that of the dirty case when
  $C/N \rightarrow 0$, since
  the first term is then dominant. Thus (\ref{6.17}) gives the correction
to the
dirty limit, and it is valid for $C \sim N$.

 In the large $C$, $C >> N$, it is easy to recover the clean case
 from the expression  (\ref{6.16}).
 The saddle-point is given by $t = - i\mu $ and $v = \lambda$. Putting these
 values in (\ref{6.16}), and taking $\mu \rightarrow \lambda$,
 we obtain immediately
 \bq
 \rho(\lambda) = {N\over{2 C}} {\rm sin}^2 ({2 N \lambda\over{C}})
 \eq

 Thus, (\ref{6.16}) gives the result both for the dirty case and for
 the clean case when one varies the parameter $C$.

\section{DISCUSSION}

 We have discussed a random matrix theory for the
 energy levels inside a superconductor vortex
 and investigated the crossover from the clean case to the dirty
 case.

 The technique involves a transposition of earlier results for the
Hermitian case to the
 symplectic group $Sp(N)$. An extension of
the Itzykson-Zuber integral is presented (in an appendix). In the
symplectic case
we have to consider an external source matrix,
which describes the spectrum of energies in the absence of impurities,
 whose eigenvalues are of the form
 $a = (a_1,...,a_N,-a_1, ...,-a_N)$.
 In the symplectic case the measure $\Delta (a)= \prod (a_i^2 - a_j^2)
\prod a_i$ replaces
the Vandermonde determinant of the unitary case.
 Indeed, in the superconductor vortex,   the eigenvalues appear in opposite
 pairs, i.e. with an exact  particle-hole symmetry,
 as seen in the Andreev reflexion.
 We have then discussed  generalized integral formulae related to the $Sp(N)$
  group. In an appendix we have  investigated  hypergeometric functions of
  matrix argument both for the unitary and symplectic groups.
  This result may be useful for  non-Gaussian random distributions.

This random matrix theory is  phenomenological. One
phenomenological parameter, which can be viewed as the intensity of the
disorder, is
sufficient to describe  the
distribution functions of the random matrices. We have fixed this
disorder parameter to one, which is why it does not appear explicitly in
(\ref{4.2}).
Instead for the external source matrix $A$, we have introduced a parameter
$C$ in
(\ref{5.16}). We find the a universal formula for the density
of states and the correlation functions for the different
regimes of this parameter: a clean case $C > N$, a dirty case $C < N$,
and a crossover region $C \sim N$.
Nevertherless the identification of this parameter $C$ with  microscopic
parameters, such as the strength of the impurity potential $V_i$ and the
number of impurities
$N_i$ in (\ref{2.7}), requires  a microscopic
study which is outside the scope of this work.

\acknowledgements
This work was supported by the CREST of JST.  S. H. thanks a Grant-in-Aid
for Scientific Research by the Ministry of Education, Science and Culture.
A. L. thanks a Grant NSF DMR-9812340. S. H. and A. L. thanks
ICTP in Trieste where this work was started.

%***********************
\newpage
\setcounter{equation}{0}
\renewcommand{\theequation}{A.\arabic
{equation}}
{\bf Appendix A: {Harish-Chandra formula for the unitary, orthogonal and
symplectic group}}
\vskip 5mm

Let $G$ be a compact connected Lie group.
The Harish-Chandra integral formula is
\cite{Harish-Chandra,AltItzykson,GrossRichards1}
\bq\label{A1}
\int_G e^{<{\rm Ad}(g)\cdot a|b>} dg = {\sum_{w\in W} ({\rm det} w)
e^{<w\cdot a|b>}
\over{\Delta(a)\Delta(b)}}
\eq
where $a, b\in h$; $h$ is a Lie algebra, and for any $H\in h$,
\bq\label{A2}
\Delta(H) = \prod_{\alpha\in \Delta_{+}} \alpha(H)
\eq
$\Delta_{+}$ is  the collection of positive roots,
 $W$ is the finite reflection group, called the Weyl (or Coxeter) group
and $h$ is the Cartan subalgebra of the Lie algebra of the group.
In the general classification theory, the irreducible finite
refelection groups are categorized as belonging to various types,
$A_n,B_n,C_n,D_n,....$, which are associated with certain compact Lie
groups.
In (\ref{HC}), ${\rm det} (w)$ is simply $\pm 1$, since each $w\in W$
is an orthogonal transformation.

Let us illustrate this result in the simplest case of the unitary group.
The Lie algebra of $SU(n)$ ,  $u = su(n)$,
consists  $n\times n$ complex skew-Hermitian traceless matrices.
The complexification of $u$ is $g = sl(n,C)$, the Lie algebra of all
$n\times n$ complex matrices with zero trace.
The Cartan subalgebra $h$ of $g$ consists of  diagonal $n\times n$
complex matrices $H = diag(h_1,...,h_n)$ such that $h_1 + ....
+h_n = 0$.
We define the linear functional $e_j$ on $h$ by $e_j(H) = h_j$,
and the $n\times n$ matrix $E_{jk}$   which consists of 1 in the (j,k)th
position and 0 elsewhere. The  linear functional $\alpha = e_j - e_k,j\neq
k$ is
a root of $g$ with respect to $h$, i.e.
$\Delta = [e_j - e_k: 1 \leq j\neq k\leq n]$.
One then verifies that (\ref{A1}) reduces the Itzykson-Zuber formula
(\ref{HC}) in the unitary case.

The orthogonal group $O(N)$ consists of real $N\times N$ matrices $u$ such that
$u u^{t} = 1$. The Harish-Chandra formula applies to  compact connected
groups $G$
and we thus restrict ourselves to the special orthogonal
subgroup $SO(N)$, of orthogonal matrices with determinant one. The Lie
algebra  $g = so(N)$
of $ G = SO(N)$ consists of all
$N\times N$ real skew-symmetric traceless matrices.
We need to consider separately the even case $N = 2n$ and the odd one $N =
2n + 1$.

For $SO(2n)$, the Cartan subalgebra $h$ is
the complex Lie algebra of all $2n \times 2n$ complex block-diagonal
matrices of the form \cite{KNAPP}
\bq\label{A3}
H = \left(\matrix{0 & h_1 & & & & & & &\cr
                  -h_1&0  & & & & & & &\cr
                      &   & 0 & h_2& & & & &\cr
                      &   & -h_2& 0& & & & &\cr
                      &   &     &  &\cdot & & & &\cr
                      &   &     &  & &\cdot & & &\cr
                      &   &     &  &  &   &0&h_n&\cr
                      &   &     &  &  &   &-h_n&0&}\right)
\eq

This matrix $H$ is written as the direct sum of $v$'s defined as
\bq\label{A4}
v = \left(\matrix{0&1\cr
                  -1&0}\right)
\eq
\bq\label{A5}
H = h_1 v \oplus h_2 v \oplus \cdots \oplus h_n v
\eq
Let $[e_1,...,e_n]$ be the standard basis for $R^n$.
A root system for $SO(2n)$ is $\Delta = [\pm e_j \pm e_k:
1\leq j \le k \leq n]$, which is a root system of type $D_n$.
For $\alpha = e_j \pm e_k\in \Delta_{+}$ and the
Cartan subalgebra $H$, we have $\alpha(H) = h_j \pm h_k$
and
\bq\label{A6}
V(H) = \prod_{1\leq j < k \leq n} (h_j^2 - h_k^2).
\eq
A fundamental Weyl chamber is defined as $S = [(h_1,...,h_n)\in R^n:h_1
>\cdots > h_{n-1} > |h_n|]$.
The group $G(n)$ of permutations $w$ of the set
$[-n,...,-1,1,...,n]$ restricted to  $w(-j)$ = - $w(j)$,
acts on the set of $[h_{-n},...,h_{-1},h_1,...,h_n]$
as
\bq\label{A7}
w\cdot (h_1,...,h_n) = (h_{w(1)},...,h_{w(n)})
\eq
where we denote $h_{-j} = - h_j, j = 1,...,n$.
The Weyl group $W$ consists of  $W$ = [permutations and even number of
sign changes of $[e_1,
...,e_n]$]; thus  $|W| = 2^{n-1}n!$.
For $H$, we have
\bq\label{A8}
w\cdot H = h_{w(1)}v\oplus h_{w(2)} v\oplus \cdots \oplus h_{w(n)} v
\eq
Therefore the Harish-Chandra formula (\ref{A1}) gives for $a = a_1 v \oplus
a_2 v\oplus \cdots
\oplus a_n v$ and $b = b_1 v\oplus b_2 v\oplus \cdots \oplus b_n v$
\bq\label{A9}
\int_{SO(2n)} e^{{\rm tr} ( g a g^{-1} b)} dg
= C_{SG(n)} {\sum_{w\in SG(n)} ({\rm det} w) \exp ( 2 \sum_{j=1}^n
w(a_j) b_j)\over{
\prod_{1\leq j < k \leq n}(a_j^2 -a_k^2)(b_j^2 - b_k^2)}}
\eq
with $C_{SG(n)} = ( 2n - 1)! \prod_{j=1}^{2n -1} (2 j - 1)!$.
%************%
\vskip 2mm
For $SO(2n + 1)$, the matrix $H$ is
\bq\label{A10}
 H = h_1 v \oplus h_2 v\oplus \cdots \oplus h_{n} v\oplus 0
\eq
and the root system for $SO(2n + 1)$ is
\bq\label{A11}
\Delta = [ \pm e_j \pm e_k: 1\leq j \le k \leq n ]\cup[
\pm e_j:1\leq j\leq n]
\eq
Thus for $\alpha\in \Delta_{+}$ and $H$, we have
\begin{eqnarray}\label{A12}
\alpha(H) =& & h_j\pm h_k \hskip 24mm if \hskip 5mm\alpha = e_j \pm e_k\cr
           & & h_j  \hskip 35mm if \hskip 5mm\alpha = e_j
\end{eqnarray}
and
\bq\label{A13}
   V(H) = \prod_{1\leq j < k \leq n }( h_j^2 - h_k^2) \prod_{j=1}^n
h_j
\eq
The weyl group of $SO(2n + 1)$ is $W$ =[ permutations and sign changes of
$[e_1,
...,e_n]$]. $|W| = 2^n n!$.

 The action of $G(n)$ on the Lie algebra $h$ is
\bq\label{A14}
w\cdot H = h_{w(1)} v\oplus h_{w(2)} v\oplus \cdots
\oplus h_{w(n)} v\oplus 0
\eq
The Harish-Chandra formula (\ref{HC}) becomes for
$a = a_1 v\oplus \cdots \oplus a_n v\oplus 0$, $b = b_1 v\oplus \cdots
\oplus b_n v \oplus 0$,
\bq\label{A15}
\int_{SO(2n + 1)} e^{{\rm tr}(g a g^{-1} b)} dg
= C_{G(n)} {\sum_{w\in G(n)} ({\rm det} w) \exp(2 \sum_{j=1}^n w(a_j) b_j)
\over{\prod_{ 1\leq j \le k \leq n}(a_j^2 - a_k^2)(b_j^2 - b_k^2)
\prod_{j=1}^n a_j b_j}}
\eq
where $C_{G(n)} = \prod_{j=1}^n(2j - 1)!\prod_{j=2n}^{4n - 1} j!
$.
%*********%
\vskip 2mm
For the case of the symplectic group $Sp(N)$, we have
\bq\label{A16}
H = \left(\matrix{h_1& & & & & & &\cr
                     & \cdot & & & & & &\cr
                     & & h_n& & & & & \cr
                     & & & - h_1& & & &\cr
                     & & & & \cdot & & &\cr
                     & & & & & -h_n& &}\right)
\eq
and $e_j(H) = h_j$.
A root system for $Sp(N)$ is
\bq\label{A17}
\Delta = [\pm e_j \pm e_k: 1\leq j \le k \leq n ]\cup[
\pm 2 e_j:1\leq j\leq n]
\eq
The Weyl group for the Sp(N) algebra is $W$ = [permutations and
sign changes of $[e_1,...,e_n]$]. $|W| = 2^{n}n!$.
\vskip 8mm

%**************************

\setcounter{equation}{0}
\renewcommand{\theequation}{B.\arabic
{equation}}
{\bf Appendix B: {Generalization of the Itzykson-Zuber formula}} \vskip 5mm

The Itzykson-Zuber formula for  Hermitian matrices $a$ and $b$ is again
\bq\label{B1}
 \int_{U(n)} e^{{\rm tr} (g a g^{-1} b)} dg = {{\rm det}(e^{a_ib_j})\over
 {\Delta(a)\Delta(b)}}
 \eq
 where $a_i$ and $b_i$ are eigenvalues of $a$ and $b$, respectively.
 We wish now to consider generalizations of this type of integrals for which

 \bq\label{B2}
 \int_{U(n)} \psi_n( g a g^{-1} b ) dg = {{\rm det} ( f(a_i b_j))\over
 {\Delta(a)\Delta(b)}}
 \eq
 where $\psi_n$ is a function of a matrix argument, and $f$ is a real
 function.

 The Itzykson-Zuber formula of (\ref{B1}) corresponds to  $f(x y) = \exp[ x
y]$ and $\psi_n = \exp [
 {\rm tr}(g a g^{-1} b)]$.

 We may take for $f(x y)$ an hypergeometric function $_p{\cal
F}_q(\alpha_1,...,\alpha_p;\beta_1,...,\beta_q;xy)$
  . Note that
 $f(xy) = \exp[ x y ] = {_0{\cal F}_0} ( x y )$.
 Then , the corresponding function  $\psi_n (t)$ is also an hypergeometric
 function of the matrix argument
 $_pF_q(\alpha_1 + n - 1, ..., \alpha_p + n - 1; \beta_1 + n - 1,...,
 \beta_q + n - 1; t)$.
 For example,in the  $p=1,q=0$ case, we have
 \begin{eqnarray}\label{B3}
 \int_{U(n)} {\rm det} ( 1 - a g b g^{-1} )^{-\alpha -n + 1} dg
 &=& _1F_0(\alpha; a,b)\nonumber\\
 &=& {{\rm det}(_1{\cal F}_0(\alpha; a_i b_j))\over{\Delta(a)\Delta(b)}}
 \end{eqnarray}
 where $_1{\cal F}_0(\alpha;a_ib_j) = (1 - a_ib_j)^{-\alpha}$.
 This is easily checked directly for the  $n = 2$, $g =
U(2)$ case. We represent $g$ as
\bq\label{B3a}
g = \left(\matrix{\cos \phi e^{i \theta_1}& \sin \phi e^{i\theta_2}\cr
- \sin \phi e^{i \theta_3}& \cos \phi e^{-i(\theta_1 - \theta_2
-\theta_3)}}\right)
\eq
with the measure $J = \cos \phi \sin \phi d \phi \prod_{i=1}^{3} d\theta_i$.

 More generally, the formula of the integration over the $U(N)$ group
 may be written as
 \begin{eqnarray}\label{B4}
 &&_pF_q(\alpha_1,...,\alpha_p;\beta_1,...,\beta_q; a, b)\nonumber\\
 &=& \int_{U(n)} {_pF_q}(\alpha_1,...,\alpha_p;\beta_1,...,\beta_q; a g b
g^{-1})
 d g \nonumber\\
 &=& C {{\rm det}(_p{\cal F}_q(\alpha_1 - n + 1,...,
 \alpha_p -n + 1; \beta_1 - n + 1,...,\beta_q - n + 1; a_i b_j))
 \over{\Delta(a)\Delta(b)}}\nonumber\\
 \end{eqnarray}

 This derivation of (\ref{B4}) proceeds by induction\cite{Gross}.
 First, we introduce the zonal polynomials $Z_m(A)$ \cite{James} which are
homogeneous polynomials of degree m, which are symmetric functions of the
$n$ eigenvalues of the matrix $A$.
From their definition they have the simple property that
\bq\label{B7}
\int_{G} Z_m( g A g^{-1} B) dg = {Z_m(A) Z_m(B)\over{Z_m(1)}}
\eq
in which the integral runs over the elements of a compact Lie group $G$
with a Haar measure normalized to one.
The coefficients of these polynomials are group-dependent, but they can be
constructed
from this property inductively. The polynomials  are thus expressed as
decompositions of
products of $ Tr(A^m_1)Tr(A^m_2)\cdots Tr(A^m_n)$, with $|m| = m_1 + \cdots
+ m_n$, and the $m_j$
are the partitions of m characterizing a   Young tableau. It then follows that
 \bq\label{B5}
 ({\rm tr} A)^{k} = \sum_{|m| = k} Z_m(A)
\eq
in which the sum runs over all the Young tableaux with k boxes.
We may now take for a generating function the
hypergeometric function with matrix argument  defined as
\bq\label{B6}
 _pF_q(\alpha_1,...,\alpha_p;\beta_1,...,\beta_q;t) =
 \sum_{k=0}^{\infty} {1\over{k!}} \sum_{|m| = k} {[\alpha_1]_m\cdots
 [\alpha_p]_m\over{[\beta_1]_m\cdots [\beta_q]_m}}
 Z_m(t).
\eq
Then, one has
 \begin{eqnarray}\label{B8}
 _pF_q(\alpha_1,...,\alpha_p;\beta_1,...,\beta_q;a,b) &=&
 \int_{G}{_p}F_q(\alpha_1,...,\alpha_p;\beta_1,...,\beta_q; gag^{-1}b) dg
 \nonumber\\
 &=&
 \sum_{k=0}^{\infty} {1\over{k!}} \sum_{|m| = k} {[\alpha_1]_m\cdots
 [\alpha_p]_m\over{[\beta_1]_m\cdots [\beta_q]_m}}
 {Z_m(a)Z_m(b)\over{Z_m(1)}}\nonumber\\
 \end{eqnarray}
 where
 \bq\label{B9}
 [\alpha]_m = \prod_{j=1}^n ( \alpha - j + 1)_{m_j}
 \eq
 \bq\label{B10}
  (\alpha)_k = \alpha(\alpha + 1) \cdots (\alpha + k - 1)
  \eq
  In the case of the unitary group $G=U(n)$ the zonal polynomial $Z_m(A)$
is a  Schur function,
namely it is given by the Weyl formula for the characters of the
representations of G
  \bq\label{BSchur}
  Z_m(A) = \frac {\det  (a_i^{m_j+n-j})}{\det  (a_i^{n-j})}.
  \eq
  The Euler integral gives
  \bq\label{B11}
  Z_m(A) = {\Gamma_n(\beta)\over{\Gamma_n(\beta - \alpha) \Gamma_n(\alpha)}}
  {[\beta]_m\over{[\alpha]_m}} \int_{0<r<1}
  Z_m(r A) {\rm det}(r)^{\alpha - n}
  {\rm det}(1 - r)^{\beta - \alpha - n}d r
  \eq
  where the integration is over Hermitian matrices
  $r$  whose eigenvalues are between 0 and 1.
  The Gamma function $\Gamma_n(\alpha)$ is defined by
  \bq\label{B11a}
  \Gamma_n(\alpha) = \pi^{n(n-1)/2} \prod_{i=1}^n \Gamma (\alpha - i + 1)
  \eq
  Using these notations, we have established the recurrence formula,
  \begin{eqnarray}\label{B12}
   &&_{p+1}F_{q+1}(\alpha_1,...,\alpha_{p+1};\beta_1,...,\beta_{q+1};a,b)
 = {\Gamma_n(\beta_{q+1})\over{\Gamma_n(\alpha_{p+1})
 \Gamma_n(\beta_{q+1}-\alpha_{p+1})}}\nonumber\\
 &&\times 
 \int_{0<r<1} {\rm det}(r)^{\alpha_{p+1}- n}
 {\rm det}(1 - r)^{\beta_{q+1} - \alpha_{p+1} - n}
 {_p}F_q(\alpha_1,..,\alpha_p;\beta_1,..,\beta_q;ra,b) dr
 \nonumber\\
 \end{eqnarray}\label{B13}
 which proves (\ref{B4}) inductively.

  For the case of the confluent and Gaussian hypergeometric
  functions, $_1F_1$ and $_2F_1$, these formulae reduce to
  \bq\label{B14}
  _1F_1(\alpha;\beta;t) =
  {\Gamma_n(\beta)\over{\Gamma_n(\alpha)\Gamma_n(\beta - \alpha)}}
  \int_{0<r<1} dr e^{{\rm tr}(r t)} {\rm det}(r)^{\alpha - n}
  {\rm det}( 1 - r)^{\beta - \alpha - n}
  \eq
  \begin{eqnarray}\label{B15}
  _2F_1(\alpha,\beta;\gamma;t) &=&
  {\Gamma_n(\gamma)\over{\Gamma_n(\beta)\Gamma_n(\gamma - \beta)}}
   \int_{0<r<1} dr {\rm det}(r)^{\beta - n}
  {\rm det}( 1 - r)^{\gamma - \beta - n}\nonumber\\
  &&\times {\rm det}(1 - r t)^{-\alpha}
  \end{eqnarray}

   For the symplectic group $Sp(n)$, we have similar
  formulae with hypergeometric functions. In the case $p=0,q=0$,
and $p=1,q=0$, which is similar to (\ref{B1}) and (\ref{B3}),
\begin{eqnarray}\label{B15a}
{_0}F_0(a,b) &=& \int_{Sp(n)} e^{{\rm tr}(g a g^{-1} b)} dg\nonumber\\
&=& {{\rm det} [ 2 {\sinh}(2 a_i b_j) ]\over{\Delta(a)
\Delta(b)}}
\end{eqnarray}
\begin{eqnarray}\label{B16}
  {_1}F_{0}(\alpha;a,b) &=&
  \int_{Sp(n)} {\rm det} ( 1 - a g b g^{-1})^{-\alpha - 2 n + 1}
 {\rm d}g
  \nonumber\\
  &=& {{\rm det}[({1\over{1 -a_ib_j}})^{2 \alpha}
  - ({1\over{1 + a_ib_j}})^{2 \alpha}]\over{\Delta(a)\Delta(b)}}
  \end{eqnarray}
  where $\Delta(a) = \prod (a_i^2 - a_j^2) \prod a_i$.
These formula are easily checked  for the simplest
case, $n = 1$, $Sp(1)$, which is isomorphic to $SU(2)$,
with
\bq\label{B17}
g = \left(\matrix{ \cos \phi e^{i \theta_1}& \sin \phi e^{i\theta_2}\cr
- \sin \phi e^{- i \theta_2}& \cos \phi e^{- i \theta_1}}\right)
\eq
It may be useful to notice that these results for $Sp(1)$ may also be
derived immediately from the $U(2)$ case by
setting $a = a_1 = - a_2$, and $b = b_1 = - b_2$ in (\ref{B1})
and (\ref{B3}). Indeed the Lie group $g$ of $Sp(1)$ in (\ref{B17})
is derived from (\ref{B3a}) with the condition $\theta_3 = - \theta_2$.
The denominator $\Delta(a)$ for $Sp(n)$ is
also derived from the $U(n)$ case with the condition on the
eigenvalues $ a = {\rm diag}(a_1, \cdots, a_n, - a_1, \cdots,
- a_n)$.
This equivalence between $U(n)$ and $Sp(n)$ holds for $n \neq 1$.
Since the eigenvalues appear here in pairs
$(a_i,- a_i)$, we have to replace $n$ by $2 n$ in the hypergeometric
relations as shown in (\ref{B16}).
For a general hypergeometric function, we have for the $Sp(n)$ case,
\begin{eqnarray}\label{B18}
 &&_pF_q(\alpha_1,...,\alpha_p;\beta_1,...,\beta_q; a, b)\nonumber\\
 &=& \int_{Sp(n)} {_pF_q}(\alpha_1,...,\alpha_p;\beta_1,...,\beta_q; a g
 b g^{-1})
 d g \nonumber\\
 &=& {C\over{\Delta(a)\Delta(b)}}
 {\rm det}[(_p{\cal F}_q(\alpha_1 - 2n + 1,...,
 \alpha_p -2n + 1; \beta_1 - 2n + 1,...,\nonumber\\
 && \beta_q -2 n + 1; a_i b_j))^2
\nonumber\\
&-&
(_p{\cal F}_q(\alpha_1 - 2n + 1,...,
 \alpha_p -2n + 1; \beta_1 - 2n + 1,...,\nonumber\\
 &&\beta_q - 2n + 1;  - a_i b_j))^2]
 \nonumber\\
 \end{eqnarray}


\begin{references}
\bibitem{CDM} C. Caroli, P. G. De Gennes and J. Matricon, Phys. Lett.
{\bf 9}, (1964) 307.
\bibitem{BH1} E. Br{\'e}zin and S. Hikami, Nucl. Phys. {\bf B479} (1996) 697.
\bibitem{BHZ} E. Br{\'e}zin, S. Hikami and A. Zee, Nucl. Phys. {\bf 464}
(1996) 411.
\bibitem{BH2} E. Br{\'e}zin and S. Hikami, Phys. Rev. {\bf E 55} (1997) 4067.
\bibitem{BH3} E. Br{\'e}zin and S. Hikami, Phys. Rev. {\bf E 56} 264
(1997).\bibitem{Harish-Chandra} Harish-Chandra, Proc. Nat. Acad. Sci. {\bf
42} 252
(1956).
\bibitem{Itzykson} C. Itzykson and J. -B. Zuber, J. Math. Phys. {\bf 21}
(1980) 411.
\bibitem{Duistermaat} J. J. Duistermaat and G. H. Heckman, Invent. Math.
{\bf 69} (1982) 259.
\bibitem{Kazakov} V. A. Kazakov, Nucl. Phys. {\bf B 354} (1991) 697.
\bibitem {Mehta} M.L. Mehta, Random Matrices, {\it{Academic Press}} (1991)
\bibitem{KL1} A. A. Koulakov and A. I. Larkin, Cond-mat/9810125.
\bibitem{LO} A. I. Larkin and Yu. N. Ovchinnikov, Phys. Rev. {\bf B 57} (1998)
5457
\bibitem{SKF} M. A. Skvortsov, V. E. Kravtsov and M. V. Feigel'man,
Cond-mat/9805296.
\bibitem{KL2} A. A. Koulakov and A. I. Larkin, Cond-mat/9802002.
\bibitem{AZ} A. Altland and M. R. Zirnbauer, Phys. Rev. Lett. {\bf 76}
(1996) 3420.
\bibitem{AST} A. Altland, B. D. Simon and D. Taras-Semchuk, cond-mat/9807371.
\bibitem{AltItzykson} D. Altshuler and C. Itzykson, Ann. Inst. Heri
Poincar{\'e},
{\bf 54} (1991) 1.
\bibitem{GrossRichards1} K. I. Gross and D. S. P. Richards,
J. Approx. Theory {\bf 59} (1989) 224, J. Approx. Theory {\bf 82}(1995) 60.
\bibitem{Witten} E. Witten, J. Geom. Phys. {\bf 9} (1992) 303.
\bibitem{Atiyah} M. F. Atiyah and R. Bott, Topology {\bf 23} (1984) 1.
\bibitem{HM} S. Hikami and T. Maskawa, Prog. Theor. Phys. {\bf 67}
(1982) 1038.
\bibitem{KNAPP} A. W. Knapp, {\it Representation theory of semisimple groups},
Princeton University Press, Princeton, New Jersey 1986.
\bibitem{Erdelyi} Erdelyi et al. Higher transcendental functions,
Vol.2, P.199, (1953) McGraw-Hill, New York.
\bibitem{James} A. T. James, Ann. Math. Statist. {\bf 35} (1964) 475.
\bibitem{Gross} K. I. Gross and D. S. P. Richards,
Trans. American Math. Soc. {\bf 301} (1987) 781.
\end{references}
\end{document}